\documentclass[aps,prl,preprint,groupedaddress]{revtex4}
\usepackage{graphicx}

\begin{document}

\title{Comment on "Direct evidence for strong crossover of collective excitations and fast sound in the
supercritical state"
}

\author{T. Bryk$^{1,2}$, I. Mryglod$^{1}$
}

\affiliation{
$^1$Institute for Condensed Matter Physics of the
National Academy of Sciences of Ukraine,
1 Svientsitskii Street, UA-79011 Lviv, Ukraine\\
$^2$ Institute of Applied Mathematics and
Fundamental Sciences, Lviv Polytechnic National University, UA-79013
Lviv, Ukraine\\
}

\date{\today}

\begin{abstract}
We comment on three incorrect claims in the paper by Fomin {\it et
al} (arXiv:1507.06094) concerning the generalized hydrodynamic
methodology and positive sound dispersion in fluids.
\end{abstract}

\pacs{61.20.Ja,61.20.Lc, 62.60.+v}
\keywords{generalized hydrodynamics, collective excitations, dynamics of fluids,
          positive dispersion, molecular dynamics simulations}

\maketitle

In a recent paper \cite{Fom15} the authors reported results on dispersion of
collective excitations in simple Lennard-Jones fluids obtained numerically
from molecular dynamics
simulations. Fomin {\it et al} have made several incorrect statements on a methodology
of generalized hydrodynamics known as the approach of generalized collective modes
(GCM) as well as they claimed they explained "the phenomenon of fast sound originating 
from transverse modes". We comment on the following three issues.

 \noindent {\bf (i).} The GCM methodology, historically originated
from the papers by D.Kivelson and T.Keyes \cite{Key71,Kiv72} and E.G.D.Cohen
with coworkers \cite{deS88}, is based on an extension of the hydrodynamic
set of equations by additional balance equations for non-conserved
quantities. The choice for the extended non-conserved dynamic variables via
the time derivatives of the hydrodynamic ones was very obvious because of
the "orthogonality" condition for equal-time static averages
$$
\langle A(-k)|{\dot A}(k) \rangle \equiv 0
$$
in equilibrium. Here the overdot means the time derivative of the dynamic variable
$A(k,t)$, where $k$ is wave number.
The extended set of $N_v$ dynamic variables allows to write down the obtained system
of equations in a matrix form of the
generalized Langevin equation for the $N_v\times N_v$ matrix of Laplace-transformed
time correlation functions
\begin{equation}\label{langev}
\tilde{\bf F}(k,z) =
\left[ z{\bf I} - {i{\bf \Omega}(k)} + \tilde{\bf M}(k,z) \right]^{-1}
{\bf F}(k,t=0),
\end{equation}
where the $N_v\times N_v$ matrices ${\bf I}$, ${\bf \Omega}(k)$, $\tilde{\bf M}(k,z)$
are the unity matrix, frequency matrix and matrix of memory functions, respectively \cite{Han,Boo},
that proves the GCM approach being essentially the memory-function formalism. This system of
$N_v$ equations can be strictly derived by the method of non-equilibrium statistical operator 
\cite{Mry98}.

In Markovian approximation the $N_v\times N_v$ generalized hydrodynamic
matrix \cite{Mry95} reads
\begin{equation} \label{tk}
{\bf T}(k) = -i{\bf \Omega}(k) + \tilde{\bf M}(k,z=0)~ \equiv {\bf
F}(k,t=0) \left[\tilde{\bf F}(k,z=0)\right]^{-1}~.
\end{equation}
and allows to control the Markov approximation for the highest-order memory
function by the order of the time derivatives in the extended dynamic
variables, see \cite{Mry97}. A basis set of $s+1$ dynamic variables which
contains a sequence of the time derivatives of hydrodynamic variable
$A(k,t)$ up to the $s$-th order when applied for theoretical description of
a time correlation function $F_{AA}(k,t)$ is formally equivalent to the
$s+1$ step in Mori's continued fraction expansion \cite{Mor65} for its
Laplace-transform ${\tilde F}_{AA}(k,z)$ with all the corresponding sum
rules satisfied. It is easily to check by straightforward algebra using the
eigenvalues of the generalized hydrodynamic matrix ${\bf T}(k)$, that the
sum rules (frequency moments) are exactly satisfied for the first $s+1$
frequency moments of ${\tilde F}_{AA}(k,z)$ if the basis set of GCM
methodology contains a sequence of time derivatives of hydrodynamic variable
$A(k,t)$ up to the $s$-th order.

Hence, it is not clear how Fomin {\it et al} came to a claim
 (quotation  from \cite{Fom15})  "First, the number of relevant variables
in this approach is guessed [27]. Not based on a fundamental principle the method is ad hoc
and is not generally applicable". This gives evidence that the authors of \cite{Fom15} do
not understand
the role of extended (non-hydrodynamic) dynamic variables and how they are chosen.
In paper \cite{Bry01} (referenced as [27] in \cite{Fom15})
a sequence of time derivatives of hydrodynamic variables was taken up to the third order
providing the precision
for description of the density-density time correlation functions within the highest by date
fourth-order memory function, that allowed to check the Markovian approximation and convergence
of the results with increasing order of memory functions.

In another quotation from \cite{Fom15} the authors claim that the GCM approach "is
internally inconsistent and violates the sum rules because the odd moments are non-zero
in the method employing exponentials for the correlators". Fomin {\it et al} refer
for this claim to a paper \cite{Lee83}, in which the velocity autocorrelation function is analyzed.
It seems Fomin {\it et al} do not see the difference between the
autocorrelation function of a non-conserved quantity (single-particle
velocity, or stress \cite{Eva80} etc) which decays with long tail $\sim
t^{-{3/2}}$ and
 hydrodynamic time correlation functions of conserved quantities,
like density-density ones which pretty well are described by exponentials
(see \cite{Bar14} for detailed
analysis).
Concerning the sum rules fulfilled - the GCM approach provides even higher number
of the sum rules than the regular memory function approach (viscoelastic model with two relaxation
times), because it provides simultaneously
at least first three frequency sum rules for the energy-energy correlations and first four
frequency rules for the density-energy correlations. As it was mentioned above the hierarchy of
memory functions provided by the time derivatives of hydrodynamic variables makes the GCM approach
very flexible in fulfillment of the exact sum rules up to any desired order. Namely a requirement
to obtain the desired level of fulfilled sum rules for the analytical GCM representation of the 
time correlation function of interest (like density-density or energy-energy ones) leads to the
unambiguous choice of extended dynamic variables.

\begin{figure}
\includegraphics[width=0.35\textwidth,height=0.20\textheight]{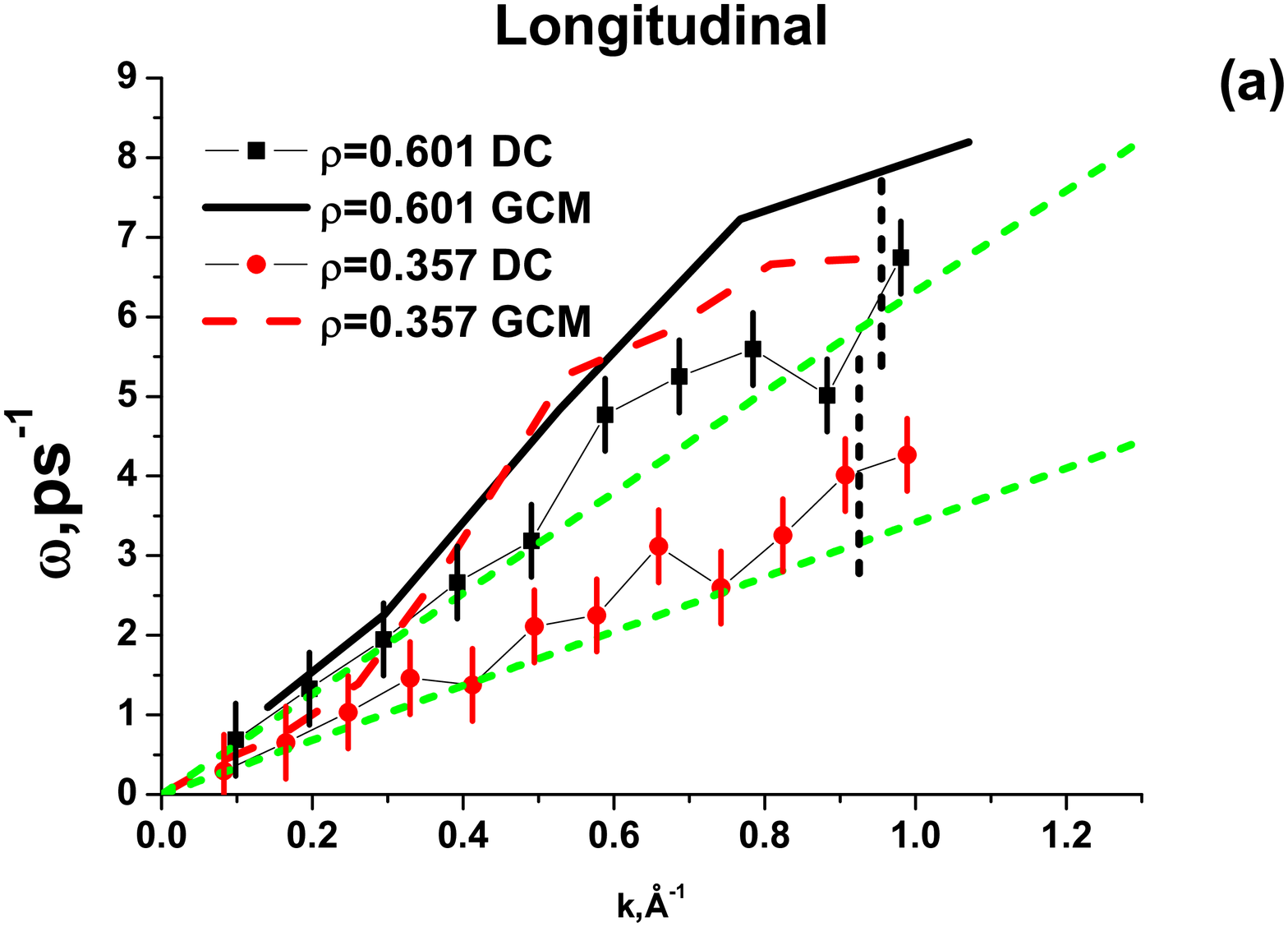}%
\includegraphics[width=0.35\textwidth,height=0.20\textheight]{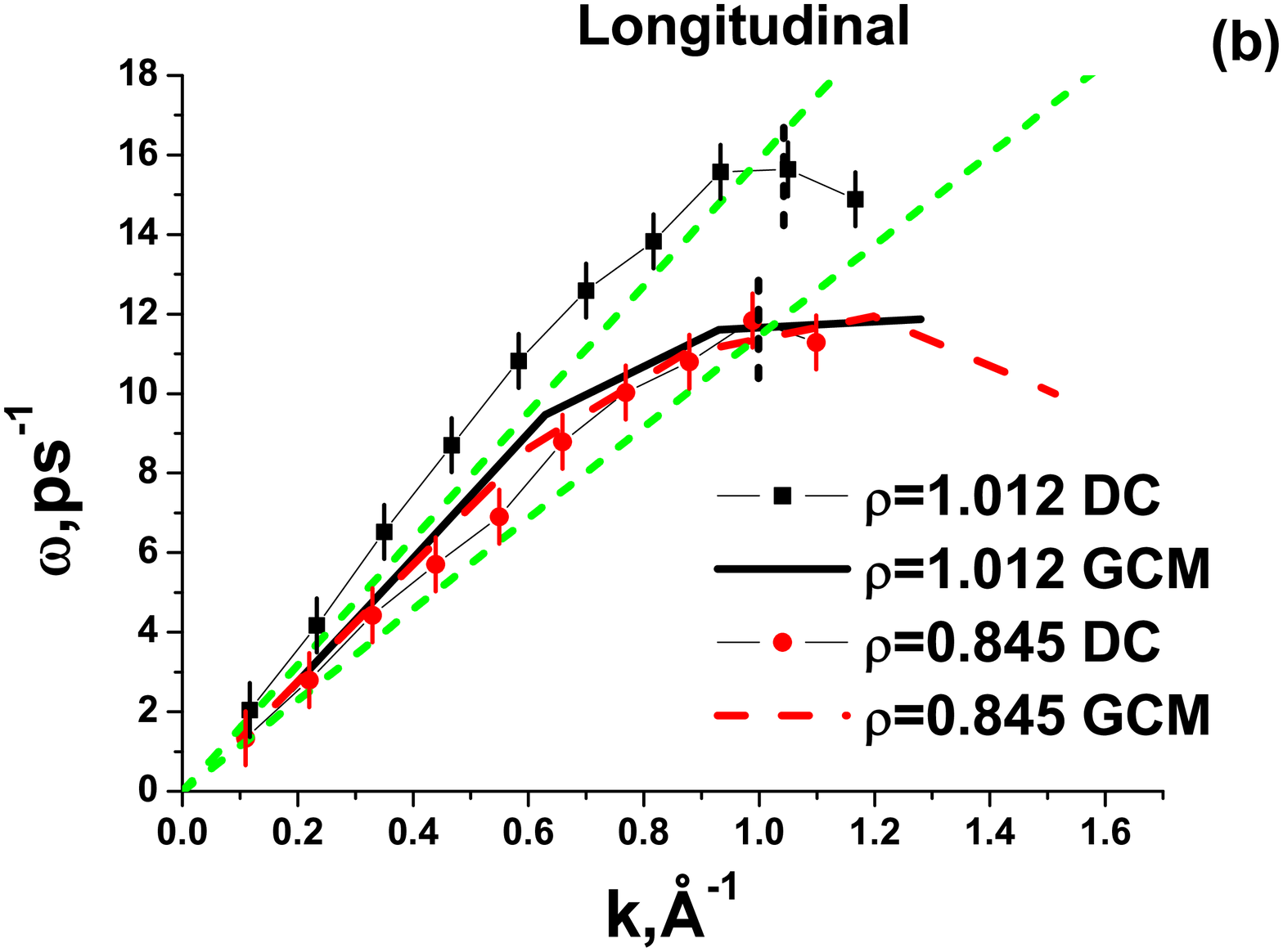}%
\includegraphics[width=0.3\textwidth,height=0.25\textheight]{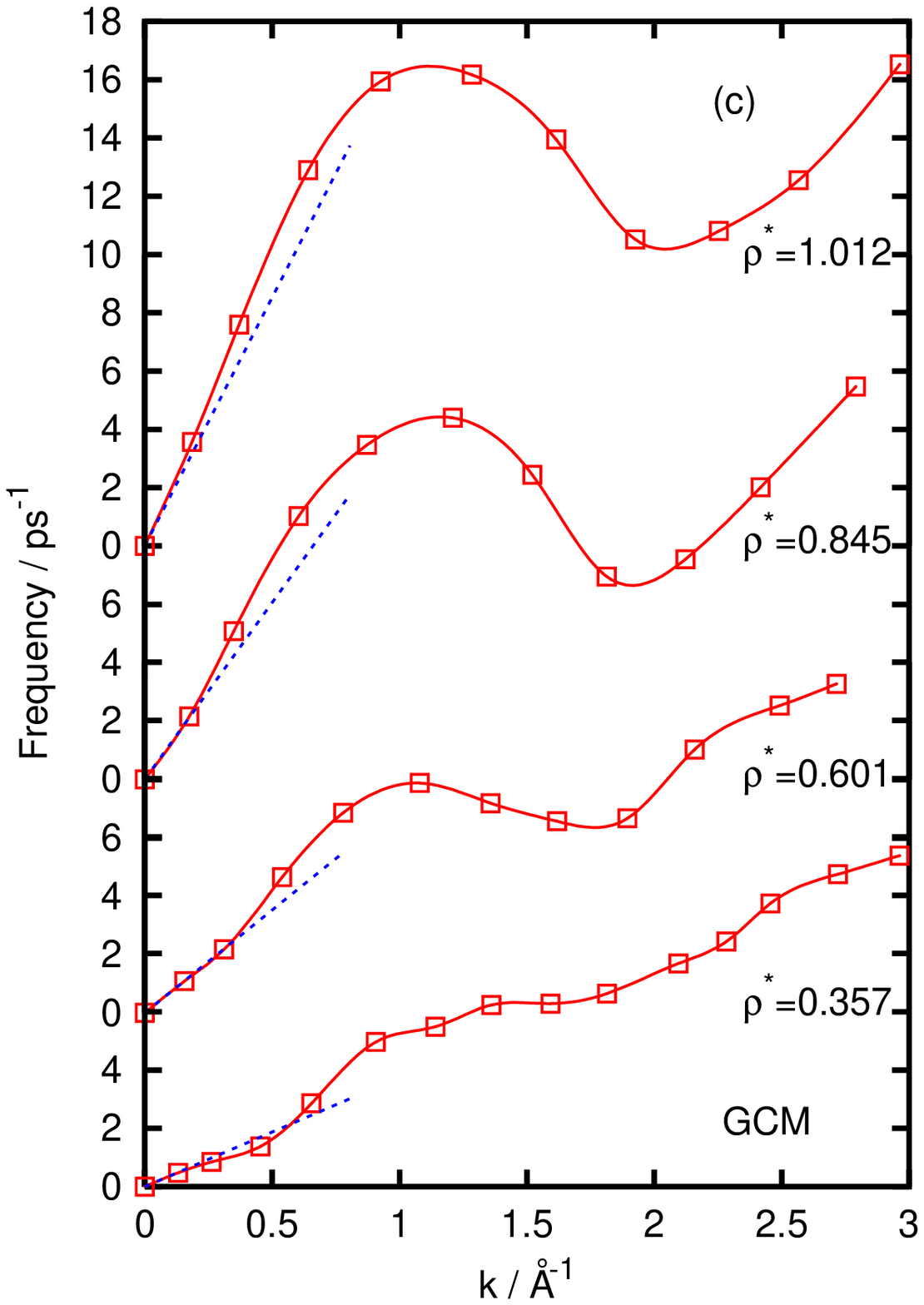}%
\caption{Dispersion curves for longitudinal collective excitations in
supercritical Ar at T$^*$=1.71 reported in \cite{Fom15} for four densities
(a,b) including the dispersions ascribed in \cite{Fom15} to the GCM results
and valid GCM eigenvalues reported at the same thermodynamic points in
\cite{Bry10} (c).
} \label{fig1}
\end{figure}

\noindent {\bf (ii).} It seems there is not only a lack of understanding of
the generalized hydrodyamic methodology by Fomin {\it et al}. In
\cite{Fom15} it was claimed that the GCM methodology numerically leads to
wrong results. The authors of \cite{Fom15} compared their estimated
dispersion curves from peak positions of the longitudinal current spectral
function for supercritical Ar at reduced temperature T$^*$=1.71 and four
densities with the GCM results taken as it is claimed from Ref.\cite{Bry10}
- see Figs.\ref{fig1}a,b. Based on these figures it was concluded in
\cite{Fom15} that "the dispersion curves derived in the GCM method stay
nearly the same at different densities, the result not expected on physical
grounds". In Fig.\ref{fig1}c we reproduce the dispersion curves for the four
densities of supercritical Ar at T$^*$=1.71 published in \cite{Bry10} and
the readers can easily compare the data and estimate how the dispersions
depend on density. It is not clear why the authors of \cite{Fom15} try to
publish obvious fake ascribed as the GCM results.

\begin{figure}
\includegraphics[width=0.5\textwidth,height=0.25\textheight]{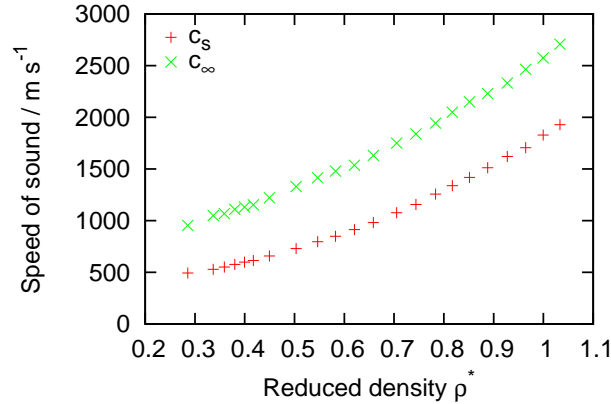}%
\caption{Adiabatic and high-frequency speeds of sound for supercritical Ar
at T=363K in a wide range of densities, obtained in molecular dynamics simulations
\cite{Gor13}. The methodology of calculations of $c_s$ and $c_{\infty}$ was the same as
described in \cite{Bry10}.
} \label{fig2}
\end{figure}

\noindent {\bf (iii).}  Fomin {\it et al} claimed that they have explained
the positive sound dispersion in liquids as "originating from transverse
modes in the supercritical state below the Frenkel line". The PSD in fluids
consists in bending up the dispersion of longitudinal collective excitations
from the linear hydrodynamic dispersion law $\omega_{hyd}(k)=c_sk$ towards
higher frequencies because of the viscoelastic effects. Here the adiabatic
speed of sound $c_s$ is defined via macroscopic quantities as
\begin{equation}\label{cs}
c_s=\sqrt{\frac{\gamma}{\rho\chi_T}}\equiv\sqrt{\frac{B_s}{\rho}}~,
\end{equation}
where $\chi_T$ is isothermal compressibility, $\gamma$ - the ratio of specific
heats, $B_s$ is the adiabatic bulk modulus, and $\rho$ is mass density.
The viscoelasticity of fluids is the consequence of two different regimes
in dynamics of fluids which are governed by different stress tensors: the viscous
one on macroscopic scale and the elastic one on atomic-resolution scale.
The viscous stress tensor depends only on the velocity field in fluid, while
the elastic stress tensor is defined by microscopic forces acting between
atomistic particles \cite{Cop75,Han,Boo}. In order to discriminate between the viscous
and elastic regimes the elastic moduli are marked in a standard way as the high-frequency ones:
the high-frequency bulk $B_{\infty}$ and shear $G_{\infty}$ moduli \cite{Boo,Han}. Corresponding
high-frequency speed of sound, which would have the idealized {\it non-damped} long-wavelength
propagating longitudinal modes in elastic regime, is
\begin{equation}\label{cinf}
c_{\infty}=\sqrt{\frac{B_{\infty}+4/3 G_{\infty}}{\rho}}~,
\end{equation}
and both speeds of sound, $c_s$ and $c_{\infty}$ are well defined in the whole range of 
densities of fluid, as
it is shown
in Fig.\ref{fig2} on example of supercritical Ar at 363~K. It is possible to define
also in analogy the high-frequency speed of propagation of idealized {\it non-damped} 
long-wavelength transverse modes
\begin{equation}\label{cinfT}
c^T_{\infty}=\sqrt{\frac{G_{\infty}}{\rho}}~,
\end{equation}
that does not make much sense because the real long-wavelength transverse acoustic modes
in fluids cannot propagate and the linear dispersion for long-wavelength transverse
excitations with propagation speed $c^T_{\infty}$ is not valid
for fluids. This is the clear difference between the idealized {\it non-damped}
(sometimes called as "bare") propagating modes and collective excitations observed in
real scattering experiments or computer simulations.

In \cite{Fom15} the authors used the expression for the high-frequency
longitudinal speed of sound (\ref{cinf}) as well as the adiabatic one with
dropped indices,
$$
v_L^2=\frac{B+4/3G}{\rho}
$$
and
$$
v_{hyd}^2=\frac{B}{\rho}~,
$$
respectively, and treating $B$ as the same quantity they substituted in the
high-frequency longitudinal speed of sound in fact the bulk modulus of the
viscous regime, that is totally incorrect. It is concluded then that the
high-frequency speed of sound is the adiabatic one enhanced by the coupling
to transverse propagating modes, in their notations being \cite{Fom15}:
$$
v_L^2=v_{hyd}^2+4/3 v_T^2~,
$$
that is totally a consequence of wrong manipulation in their expression with the adiabatic
and high-frequency bulk moduli.
Note, that the quantity $(c^2_{\infty}-c^2_s)$ appears in different generalized hydrodynamic
approaches \cite{Boo,Bry10,Sco00} and is well defined for the whole range of fluid densities
(see Fig.\ref{fig2}) however nobody associated it with the transverse excitations.

In summary, we have shown that the claim on the positive sound dispersion in fluids
"originating from transverse modes in the supercritical state below
the Frenkel line" does not have any theoretical basis and was derived from the
misused expression for the high-frequency speed of sound. The comments by authors of
\cite{Fom15} on the generalized hydrodynamics and in particular on the approach of GCM
give evidence of the lack of understanding what the GCM approach is. Moreover, in \cite{Fom15}
some dispersion curves were stated as the GCM results taken from Ref.\cite{Bry10}
 which do not depend on density of fluid - we consider this as the
obvious fake. This can be easily detected by comparison with real GCM results for dispersion
of collective excitations in supercritical Ar at different densities reported
in \cite{Bry10} (see Fig.\ref{fig1}c).

\noindent


%
%

\begin{thebibliography}{12}
\bibitem{Fom15}  Yu.D. Fomin, V.N. Ryzhov, E.N. Tsiok, V.V. Brazhkin, K. Trachenko,
                 e-print arXiv:1507.06094, (2015); published 22-07-2015;
                 http://arxiv.org/abs/1507.06094
\bibitem{Key71} T. Keyes, D. Kivelson, J. Chem. Phys. {\bf 54},
                1786 (1971).
\bibitem{Kiv72} D. Kivelson, T.Keyes, J. Chem. Phys. {\bf 57},
                4599 (1972).
\bibitem{deS88} I.M. deSchepper, E.G.D. Cohen, C. Bruin, J.C.~van~Rijs,
                W. Montfrooij, and L.A.~de~Graaf, Phys. Rev. A {\bf 38},
                271 (1988).
\bibitem{Han}   J.-P. Hansen, I.R. McDonald, {\it Theory of Simple
                Liquids} (London: Academic) (1986).
\bibitem{Boo}   J.-P. Boon, S. Yip, {\it Molecular Hydrodynamics}
                (New-York: McGraw-Hill) (1980).
\bibitem{Mry98} I.M. Mryglod, Condens. Matter Phys. {\bf 1} 753 (1998).
\bibitem{Mry95} I.M. Mryglod, I.P. Omelyan, M.V.Tokarchuk, Mol. Phys.
                {\bf 84}  235 (1995).
\bibitem{Mry97} I.M. Mryglod, I.P. Omelyan, Mol. Phys.
                {\bf 92}  913 (1997).
\bibitem{Mor65} H. Mori, Progr. Theor. Phys. {\bf 57}, 767 (1965).
\bibitem{Bry01} T. Bryk, I. Mryglod, Phys. Rev. E {\bf 63}, 051202 (2001).
\bibitem{Lee83} M.H. Lee, Phys. Rev. Lett. {\bf 51}, 1227 (1983).
\bibitem{Eva80} D.J. Evans, J. Stat. Phys. {\bf 22}, 81 (1980).
\bibitem{Bar14} F. Barocchi, E. Guarini, U. Bafile, Phys. Rev. E {\bf 90}, 032106 (2014).
\bibitem{Bry10} T. Bryk, I. Mryglod, T. Scopigno, G. Ruocco, F. Gorelli,
                M. Santoro, J.Chem.Phys. {\bf{133}}, 024502 (2010).
\bibitem{Cop75} J.R.D. Copley, S.W. Lovesey, Rep. Progr. Phys., {\bf 38}, 461 (1975).
\bibitem{Gor13} F.A. Gorelli, T. Bryk, M. Krisch, G. Ruocco,
        M. Santoro, T. Scopigno, Sci. Rep. {\bf 3}, 1203 (2013).
\bibitem{Sco00} T. Scopigno, U. Balucani, G. Ruocco, F. Sette,
                J. Phys.: Condens. Matter, {\bf 12}, 8009 (2000)
%
\end{thebibliography}
\end{document}